\documentclass[twocolumn,aps,showpacs,nofootinbib,epsfig]{revtex4}
\usepackage{amssymb}
\usepackage{amsmath}
\usepackage{graphics}
\usepackage{epsfig}
\usepackage[dvips]{color}
\newcommand{\be}{\begin{equation}}
\newcommand{\ee}{\end{equation}}
\newcommand{\bea}{\begin{eqnarray}}
\newcommand{\eea}{\end{eqnarray}}
\newcommand{\beas}{\begin{eqnarray*}}
\newcommand{\eeas}{\end{eqnarray*}}

\newcommand{\nn}{\nonumber\\}

\begin{document}
\title{Heavy flavor nuclear modification factor: more baryons than mesons
  less energy loss} 
\author{Alejandro Ayala$^1$, Eleazar Cuautle$^1$, J. Magnin$^2$, Luis Manuel
  Monta\~no$^3$ and G. Toledo S\'anchez$^4$}
\affiliation{$^1$Instituto de Ciencias Nucleares, Universidad Nacional
Aut\'onoma de M\'exico, Apartado Postal 70-543, M\'exico Distrito Federal
04510, Mexico.\\ 
$^2$Centro Brasileiro de Pesquisas F\1sicas, CBPF, Rua Dr. Xavier Sigaud
150, 22290-180, Rio de Janeiro, Brazil.\\ 
$^3$ Centro de Investigaci\'on y de Estudios Avanzados del IPN, Apartado Postal
14-740, M\'exico Distrito Federal 07000, Mexico.\\
$^4$Instituto de F\1sica, Universidad Nacional Aut\'onoma de M\'exico,
Apartado Postal 20-364, M\'exico Distrito Federal 01000, Mexico.}

\begin{abstract}
The suppression of the nuclear modification factor for heavy flavor hadrons
is usually attributed to the energy loss of heavy quarks propagating in a QCD
plasma. Nevertheless it is puzzling that the suppression is as strong as
for light flavors. We show that when accounting
for the quark momentum shift associated to the opening of the
recombination/coalescence channel for hadron production in the plasma, it is
not necessary to invoke such strong energy loss. This shift is  
expressed in terms of an increase of the heavy baryon to meson ratio in
nuclear with respect to proton collisions. When this mechanism is included
along with a moderate energy loss, data from RHIC and LHC for the nuclear
modification factor of electrons coming from heavy flavor decays as well as
of charm mesons, can be reasonably described.  
\end{abstract}

\pacs{25.75.-q, 25.75.Nq}
\maketitle


The suppression of single hadron transverse spectra in nuclear collisions,
with respect to a superposition of independent proton collisions, is one of
the main results from the BNL Relativistic Heavy Ion Collider (RHIC) and the
CERN Large Hadron Collider (LHC). This suppression is quantified in terms of
the nuclear modification factor ($R_{AA}$) and one of its main features is
that heavy flavor hadrons are equally suppressed as light
hadrons~\cite{phenixstar, STAR}. Such behavior was first obtained from the
analysis of electrons from the decay of heavy flavors and later confirmed from
the analysis of charm mesons~\cite{des, Dainese}. 

When the suppression of heavy flavors is only attributed to energy
loss in the QCD medium, the above result is surprising for if the
main contribution comes from radiative processes, the {\it dead cone
effect}~\cite{deadcone} should prevent heavy quarks from losing as much
energy as light ones. This motivated the reviewing of energy loss scenarios
to incorporate contributions from collisional processes, diffusion, geometry,
as well as dynamical properties of the medium~\cite{Models}. However even
these refined scenarios do not yet provide a fully convincing
explanation for the properties of the heavy flavor $R_{AA}$.

Much less attention has been paid to the fact that a shift of the hadron
momentum in the nuclear medium can come not only from a loss of
energy but also from a momentum redistribution when the quarks from the medium
form either baryons or mesons. This is the central idea behind the
recombination/coalescence scenario as a new channel for hadron production in a
heavy-ion environment~\cite{recombination}. In average, the three quarks
forming the baryon come from lower momentum bins than the two quarks making up
a meson. Since there are more quarks with lower momenta there is a larger
chance to form baryons than mesons. Transverse flow increases the effect since
this makes the momentum distribution for heavier particles (hadrons) to fall
less steeply than for lighter ones (mesons). A direct consequence of this
momentum redistribution is an increase of the baryon to meson ratio in nuclear
with respect to proton collisions. This ratio has been measured
for a large variety of light and strange hadrons in high-energy nuclear
collisions~\cite{experimentos}. The upshot is that for intermediate transverse 
momenta, the ratio is enhanced with respect to the corresponding one in
proton collisions. Although no measurements exist for the case of heavy
flavors, there are model calculations that describe this
enhancement~\cite{onemodel, us, Pajares}. To test this scenario in
quantitative terms, we use one of these models, the Dynamical Quark
Recombination Model (DQRM)~\cite{dqrm}, to compute the heavy baryon to meson
ratio which in turn is used to compute $R_{AA}^e$ and $R_{AA}^D$. We show that
when this increase is accounted for, only a moderate energy loss is needed to
reproduce the data. 

For definitiveness, let us concentrate on describing the nuclear modification
factor for a single heavy flavor, say charm ($c$) quarks. The number of
$c$-quarks produced in nuclear ($AA$) or proton ($pp$) collisions in a given
momentum bin can be obtained from counting the corresponding number of hadrons
with $c$-quarks, namely, the number of open charm mesons ($N^D_{AA\ /\ pp}$),
charm baryons ($N^\Lambda_{AA\ /\ pp}$) and hidden charm mesons
($N^{c\bar{c}}_{AA\ /\ pp}$) 
\bea
   N^c_{AA\ /\ pp}=(N^D_{AA\ /\ pp} + N^\Lambda_{AA\ /\ pp} 
   + N^{c\bar{c}}_{AA\ /\ pp}).
\label{cAApp}
\eea
Normalizing the proton case to the average number of binary collisions
$\langle n_b\rangle $ we have that, accounting also for a possible shift in
energy $\varepsilon$, the number of produced
$c$-quarks in one and the other environments, must satisfy 
\bea
   N^c_{AA} = \varepsilon\langle n_b\rangle N^c_{pp},
\label{equality1}
\eea
that is
\bea
   (N^D_{AA} + N^\Lambda_{AA} + N^{c\bar{c}}_{AA})
   =\varepsilon\langle n_b\rangle 
   (N^D_{pp} + N^\Lambda_{pp} + N^{c\bar{c}}_{pp}).
\label{equality2}
\eea
>From Eq.~(\ref{equality2}), we can build the nuclear modification factor for
$D$-mesons in terms of the number of charm mesons and baryons in $pp$ and
$AA$ collisions and get
\bea
   R^D_{AA}&=&\frac{N^D_{AA}}{\langle n_b\rangle N^D_{pp}}\nn
   &=&
   \varepsilon \left(1 + \frac{N^\Lambda_{pp} + N^{c\bar{c}}_{pp}}{N^D_{pp}}
   \right)
   - \frac{N^\Lambda_{AA} + N^{c\bar{c}}_{AA}}{\langle n_b\rangle N^D_{pp}}.
\label{RAAD1}
\eea
The last term in Eq.~(\ref{RAAD1}) can be written as
\bea
   \frac{N^\Lambda_{AA} + N^{c\bar{c}}_{AA}}{\langle n_b\rangle N^D_{pp}}&=&
   \left(\frac{N^D_{AA}}{\langle n_b\rangle N^D_{pp}}\right)
   \left(\frac{N^\Lambda_{AA}}{N^D_{AA}}\right) + 
   \left(\frac{N^{c\bar{c}}_{AA}}{\langle n_b\rangle N^D_{pp}}\right)\nn
   &=&
   R^D_{AA}\left(\frac{N^\Lambda_{AA}}{N^D_{AA}}\right) + 
   \left(\frac{N^{c\bar{c}}_{AA}}{\langle n_b\rangle N^D_{pp}}\right).
\label{RAAD2}
\eea
Using Eq.~(\ref{RAAD2}) into Eq.~(\ref{RAAD1}), we can write
\bea
   R^D_{AA}\left(1+\frac{N^\Lambda_{AA}}{N^D_{AA}}\right)=
   \varepsilon\left(1+\frac{N^\Lambda_{pp}}{N^D_{pp}}\right) +
   \frac{N^{c\bar{c}}_{pp}}{N^D_{pp}}(\varepsilon-\eta),
\label{RAAD3}
\eea
where we have defined $\eta\equiv N^{c\bar{c}}_{AA} / \langle
n_b\rangle N^{c\bar{c}}_{AA}$. Since the ratio of hidden charm to $D$ mesons
in $pp$ collisions, $N^{c\bar{c}}_{pp} / N^D_{pp}$, is very small, to an
excellent approximation we can rewrite Eq.~(\ref{RAAD3}) as
\bea
   R^D_{AA}\simeq\varepsilon
   \left(1+\frac{N^\Lambda_{pp}}{N^D_{pp}}\right) \Big/
   \left(1+\frac{N^\Lambda_{AA}}{N^D_{AA}}\right).
\label{solRAAD}
\eea
Therefore, even in the absence of energy loss ($\varepsilon=1$) the
nuclear modification factor for $D$ mesons is smaller than one, provided the
ratio of charm baryons to open charm mesons is enhanced in $AA$ with respect
to $pp$ collisions.

The same enhancement is responsible for the suppression of the nuclear
modification factor for heavy-flavor electrons. For definitiveness, let us
again focus on electrons originating from the decay of charm
quarks. In a given momentum bin $R^e_{AA}$ can be expressed as~\cite{us2,us}
\bea
   R^e_{AA}&=&\frac{1}{\langle n_b\rangle}
   \frac{N^\Lambda_{AA}B^{\Lambda\rightarrow e} + N^D_{AA}B^{D\rightarrow e}}
        {N^\Lambda_{pp}B^{\Lambda\rightarrow e} + N^D_{pp}B^{D\rightarrow e}}\nn
   &=&\frac{1}{\langle n_b\rangle}
   \left(\frac{N^D_{AA}}{N^D_{pp}}\right)
   \left(
   \frac{B^{D\rightarrow e} + \frac{N^\Lambda_{AA}}{N^D_{AA}}
         B^{\Lambda\rightarrow e}}
        {B^{D\rightarrow e} + \frac{N^\Lambda_{pp}}{N^D_{pp}}
         B^{\Lambda\rightarrow e} }
   \right),
\label{RAAe}
\eea 
where $B^{D,\Lambda\rightarrow e}$ is the branching ratio for the decay of $D$
mesons and charm baryons into electrons, respectively. 
Using Eq.(\ref{solRAAD}), we can write Eq.~(\ref{RAAe}) as
\bea
   R^e_{AA}&=&\frac{1}{\langle n_b\rangle}
   \left(
      \frac{N^D_{AA} + N^\Lambda_{AA}}
           {N^D_{pp} + N^\Lambda_{pp}}
      \right)
   \left[
      \frac{N^D_{AA}(N^D_{pp} + N^\Lambda_{pp})}
           {N^D_{pp}(N^D_{AA} + N^\Lambda_{AA})}
      \right]\nn
   &\times&
   \left(
   \frac{1+xN^\Lambda_{AA} / N^D_{AA}}
        {1+xN^\Lambda_{pp} / N^D_{pp}}
   \right)\nn
   &=& R^D_{AA}
    \left(
   \frac{1+xN^\Lambda_{AA} / N^D_{AA}}
        {1+xN^\Lambda_{pp} / N^D_{pp}}
   \right)\nn
   &\equiv&\varepsilon \ T^e_{AA},
\label{RAAe2}
\eea
where, in order to introduce the energy loss factor $\varepsilon$, we have used
Eq.~(\ref{equality2}) ignoring the contribution from hidden 
charm mesons. Also $x=B^{\Lambda\rightarrow e} / B^{D\rightarrow e}$ and the
function $T^e_{AA}$ is given by
\bea
   T^e_{AA}&\equiv&\left[
   \left(1+\frac{N^\Lambda_{pp}}{N^D_{pp}}\right)\Big/
   \left(1+\frac{N^\Lambda_{AA}}{N^D_{AA}}\right)
   \right]\nn
   &\times&
   \left(
   \frac{1+xN^\Lambda_{AA} / N^D_{AA}}
        {1+xN^\Lambda_{pp} / N^D_{pp}}
   \right).
\label{TAA}
\eea
\begin{figure}[t!] 
{\centering
{\epsfig{file=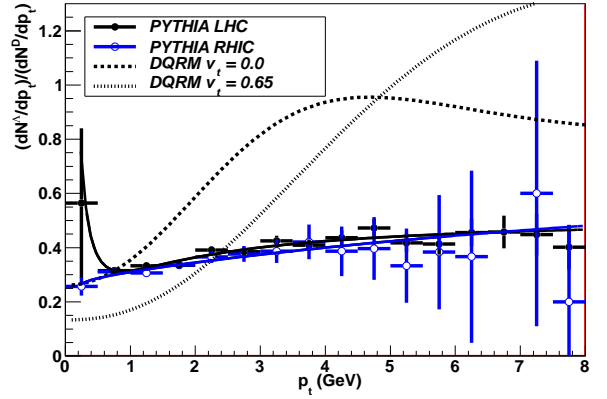, width=1\columnwidth}}
\par}
\caption{(Color on line) DQRM charm baryon to meson ratio in $AA$ compared to
the same ratio in $pp$ collisions. The DQRM curves are computed for two
transverse expansion velocities $v_t=0$ and $0.65$. For the ratio in $pp$ we
use PYTHIA simulations at $\sqrt{s_{NN}}=200$ GeV and $2.7$ TeV with $35\times
10^6$ and $15\times 10^{6}$ events, respectively. Shown are also fits to the
simulations.} 
\label{fig1}
\end{figure}
It has been shown that when $x<1$, $T^e_{AA}$ is also smaller than one
when the ratio of charm hadrons to open charm mesons is enhanced in $AA$
with respect to $pp$ collisions~\cite{us}. Therefore Eq.~(\ref{RAAe2}) states
that even in the absence of energy loss, the nuclear modification factor for
single electrons is smaller than one, provided the ratio of charm hadrons to
open charm mesons is enhanced in $AA$ with respect to $pp$ collisions and that 
electrons are more copiously produced from open charm mesons than baryons
($x<1$), which is indeed the case.

\begin{figure}[t!] 
{\centering
{\epsfig{file=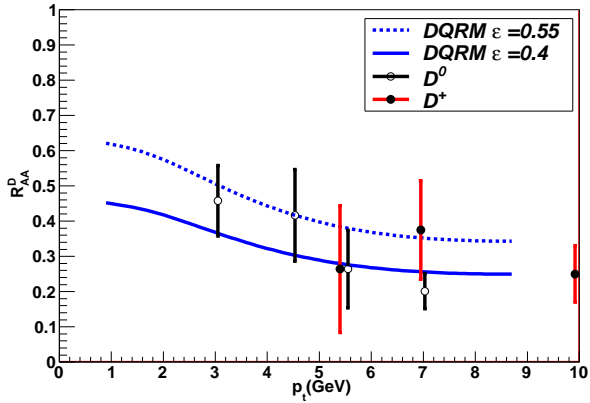, width=1\columnwidth}}
\par}
\caption{(Color on line) Nuclear modification factor for $D$ mesons compared
to ALICE data. The curves are computed using the DQRM  with $v_t=0.65$ as
appropriate for LHC energies. The charm baryon to meson ratio in $pp$ is taken
from the PYTHIA simulation described in Fig.~\ref{fig1}. For simplicity, the
energy loss parameter is taken as two constant values, $\varepsilon =0.55$
(upper curve) and $\varepsilon =0.4$ (lower curve).} 
\label{fig2}
\end{figure}

In the DQRM the probability ${\mathcal{P}}$ to recombine quarks into mesons
and baryons depends on density and temperature and thus on the proper time
$\tau$ describing the evolution of the heavy-ion reaction up to
hadronization. The evolving probability differs for hadrons made up by two and
three constituent quarks and is computed by a variational Monte Carlo
simulation. The relative population of one or the other kind 
of cluster at low densities can be fixed by combinatorial arguments (see
Refs.~\cite{us,dqrm} for details). This model is well suited to describe
baryon and meson production and its ratio at low and intermediate $p_t$. The
hadron transverse momentum distribution in central $AA$, assuming Bjorken
dynamics and transverse velocity expansion $v_t$, is given by
\bea
   \frac{dN}{p_tdp_tdy}&=&g\frac{m_t\Delta y}{4\pi}
   \frac{\rho_{\mbox{\tiny{nucl}}}}{\Delta\tau}
   \int_{\tau_0}^{\tau_f}\tau d\tau{\mathcal{P}}(\tau)\nn
   &\times&I_0(p_t\sinh\eta_t/T)e^{-m_t\cosh\eta_t/T},
\label{momdist}
\eea 
where $m_t$ is the transverse mass, $\Delta y$ the rapidity interval,
$\rho_{\mbox{\tiny{nucl}}}$ the nuclear radius, $\Delta\tau = \tau_f -\tau_0$
the proper time interval and $T$ the proper time dependent temperature
\bea
   T=T_0\left(\frac{\tau_0}{\tau}\right)^{v_s^2},
\label{temp}
\eea
with $v_s^2=1/3$. $I_0$ is a Bessel function $I$ of order zero. $v_t$ and
$\eta_t$ are related through $v_t=\tanh\eta_t$. $g$ is the degeneracy factor
that takes care of the spin degree of freedom.

Fig.~\ref{fig1} shows the DQRM charm baryon to meson ratio. We set the
masses of the charm baryon and mesons to $m^\Lambda=2.29$ GeV and $m^D=1.87$
GeV. We take the initial hadronization proper time $\tau_0=1$ fm, at an initial
temperature $T_0=175$ MeV and the final hadronization temperature $T_f=100$
MeV that, according to Eq.~(\ref{temp}), corresponds to $\tau_f=8$ fm. Shown
are the cases between $v_t=0$ and $v_t=0.65$. The figure shows also the baryon
to meson ratio in $pp$ at $\sqrt{s_{NN}}=200$ GeV and $2.7$ TeV, obtained from
PYTHIA simulations with $35\times 10^{6}$ and $15\times 10^{6}$ events,
respectively. Shown are also fits to the simulations. Notice that, as
expected, the charm baryon to meson ratio is enhanced in $AA$ with respect to
$pp$ collisions.

Fig.~\ref{fig2} shows $R^D_{AA}$ compared to ALICE data~\cite{Dainese}. The
theoretical curves are computed using Eq.~(\ref{solRAAD}) with the heavy baryon
to meson  ratio obtained in $AA$ from the DQRM with the same parameters as
before and the particular value $v_t=0.65$, which is a standard choice for the
transverse expansion velocity at LHC energies. The heavy baryon to meson ratio
in $pp$ is obtained from the PYTHIA simulation shown in Fig.~\ref{fig1} for
LHC energies. To see the effect of the energy loss parameter, for simplicity,
we take two constant values, $\varepsilon =0.55$ (upper curve) and
$\varepsilon =0.4$ (lower curve). We notice that even in this simple  
scenario, data are well described and the energy loss parameter does not need
to be as small as in the case of light flavors, which in this language means
$\varepsilon\simeq 2$, to account for the suppression in $R^D_{AA}$.

\begin{figure}[t!] 
{\centering
{\epsfig{file=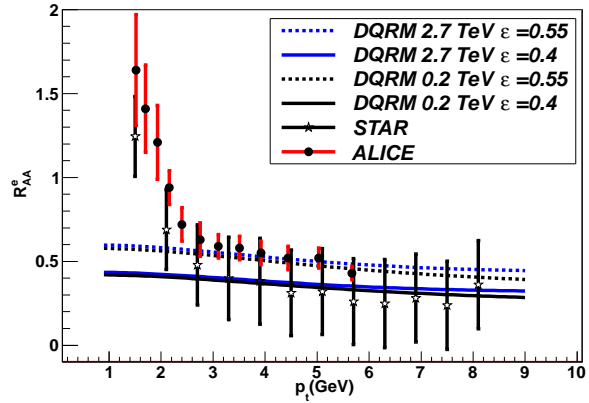, width=1\columnwidth}}
\par}
\caption{(Color on line) Nuclear modification factor for non-photonic single
electrons compared to STAR and ALICE data. The curves are computed using the
DQRM with $v_t=0.65$ for the LHC case and $v_t=0.55$ for the RHIC case. The
charm baryon to meson ratio in $pp$ is taken from PYTHIA simulations for
$\sqrt{s_{NN}}=200$ GeV and $\sqrt{s_{NN}}=2.7$ TeV for the sum of charm and
beauty hadrons. For simplicity the energy loss parameter is taken
as two constant values, $\varepsilon =0.55$ (upper curves) and $\varepsilon
=0.4$ (lower curves).}
\label{fig3}
\end{figure}

Fig.~\ref{fig3} shows $R^e_{AA}$ compared to data from STAR~\cite{STAR} and
ALICE~\cite{Dainese}. The theoretical curves are computed using
Eq.~(\ref{RAAe2}) with the heavy baryon to meson ratio obtained in $AA$ from
the DQRM and in $pp$ from the PYTHIA simulations of Fig.~\ref{fig1}. To
account for the finding that electrons from heavy flavor decays
come almost in equal proportions from the decays of charm and 
beauty hadrons for $p_t\gtrsim 5$ GeV~\cite{dataSTARe}, here we consider a
single species of heavy baryons and mesons with effective masses. We take
$m^D=3.57$ GeV, the average between the masses of the $D^0$ and the $B^0$
mesons and $m^{\Lambda}=3.95$ GeV, the average between the masses 
of the $\Lambda_c$ and the $\Lambda_b$. Also, we consider that the possible
charm and beauty mesons decaying into electrons or positrons are $D^\pm$
($B^{D^\pm\rightarrow e^\pm}=16\%$), $D^0$, $\bar{D}^0$
($B^{D^0,\ \bar{D}^0\rightarrow e^\pm}=6.53\%$),  $D^\pm_s$
($B^{D^\pm_s\rightarrow e^\pm}=8\%$) and $B^\pm$ ($B^{B^\pm\rightarrow
e^\pm}=10.8\%$), $B^0$, $\bar{B}^0$
($B^{B^0,\ \bar{B}^0\rightarrow e^\pm}=10.1\%$). The possible charm and beauty
baryons decaying into electrons or positrons are $\Lambda_c$,
$\bar{\Lambda}_c$ ($B^{\Lambda_c,\ \bar{\Lambda}_c\rightarrow e^\pm}=4.5\%$),
$\Lambda_b$ and $\bar{\Lambda}_b$ ($B^{\Lambda_b,\ \bar{\Lambda}_b\rightarrow
e^\pm}=5.35\%$) (the experimentaly reported branching ratio coreresponds to
the semileptonic decay $\Lambda_b\rightarrow \Lambda_c\ l\ \bar{\nu}_l$. Here
we consider that half of this comes from the decay into electrons). The
other parameters used for the $AA$ case are as before, with $v_t=0.55$ for the
RHIC case and $v_t=0.65$ for the LHC case. For simplicity, the energy loss
parameter is taken also as two constant values $\varepsilon=0.55$ (upper
curves) and $\varepsilon =0.4$ (lower curves). Once again, even in this simple
scenario, data are well described for $p_t\gtrsim 2$ GeV and the energy loss
parameter does not need to be as small as in the case of light flavors to
account for the supression in $R^e_{AA}$. For $p_t\lesssim 2$ GeV, the rise in
the data is usually attributed to other effects like shadowing, which is not
considered in our approach. The model curves are not significantly affected if
the effective masses are slightly varied. 

In conclusion, we have shown that when accounting for the medium's quark
momentum redistribution when these recombine/coalesce to form mesons and
baryons, the heavy flavor nuclear modification factors can be described
without the need of a large energy loss. This momentum redistribution is
encoded in the increase of the heavy baryon to meson ratio. We emphasize that
the results are valid, independent of the model as long as 
the baryon to meson ratio increases in $AA$ with respect to $pp$
collisons. This increase is expected based on general grounds, since it
represents a feature of the openning of the recombination/coalescence hadron
formation channel in $AA$ collisions. Upcoming upgrades to RHIC detectors and
to ALICE are expected to increase the capability to directly look at this
quantity and thus experimental tests of the mechanism advocated in this work
will be available in the near future.

\section*{Acknowledgments}

Support for this work has been received in part by CONACyT (Mexico) under
grant numbers 128534 and 101597 and PAPIIT-UNAM under grant numbers IN103811-3
and IN107911.

\end{document}